\documentclass[a4paper,10pt]{article}
\usepackage[utf8]{inputenc}
\usepackage[T1]{fontenc}
\usepackage{amssymb}
\usepackage{amsmath}
\usepackage{graphicx}
\usepackage{rotating}
\usepackage{parskip}
\usepackage{natbib}
\usepackage{float}
\usepackage{multicol}
\usepackage{tabularx}
\usepackage[hidelinks]{hyperref}
\usepackage[a4paper, hdivide={2.5cm,,2.5cm}, vdivide={2.5cm,,2.5cm}]{geometry}

\begin{document}


\title{Marginalization in nonlinear mixed-effects models -- \\ 
with an application to dose-response analysis}

\author{Daniel Gerhard \texttt{daniel.gerhard@canterbury.ac.nz} \\
School of Mathematics \& Statistics, \\ University of Canterbury, 
Christchurch, New Zealand \\ and \\
Christian Ritz \texttt{ritz@nexs.ku.dk} \\ Department of Nutrition, Exercise and 
Sports, Faculty of Science, \\ University of Copenhagen, Rolighedsvej 30, 1958 
Frederiksberg C, Denmark}

\date{\today}

\maketitle

\label{firstpage}

\begin{abstract}
Inference in hierarchical nonlinear models needs careful consideration about targeting parameters that have either a conditional or population-average interpretation. For the special case of mixed-effects nonlinear sigmoidal models we propose a method for the estimation of derived parameters with a marginal interpretation, but also maintaining the random effect structure of the nonlinear model, by using a combination of numerical quadrature and the delta method, integrating over the random effect distribution conditional on the estimated variance components. The difference between these marginalized estimates, generalized nonlinear least squares estimates, and conditional estimation is characterised by means of two representative case studies. The case studies consist of the estimation of effective dose levels in a human toxicology study, and the relative potency estimation for two herbicides in an agricultural field trial. Both case studies exhibit an experimental design that results in data with at least one hierarchical level of between- and within-cluster variation. A user-friendly software implementation is made available with the R package medrc, providing an automated framework for mixed-effects dose-response modelling.
\end{abstract}

\section{Introduction}

Clustered data showing nonlinear trends have recently become routine output of experiments in agriculture, biology, and toxicology (e.g., \citet{Clayton2003, Serroyen2005, Makowski2006}). For these applied disciplines parameter estimates with marginal interpretations rather than conditional interpretations would usually be relevant as such quantities provide insights and understanding of population-level behaviour in the biological systems considered. This is a very different focus as compared to pharmacokinetics and pharmacodynamics studies where interest almost solely lies in the conditional parameters \citep{Davidian1995}. Still, in practice conditional results derived from nonlinear mixed-effects models seem to be used most of the time. 

For non-normal clustered data the differences between marginal and conditional modelling are well-understood (e.g., \citet{McCulloch2008, Serroyen2009, Fitzmaurice2012, Gbur2012, Demidenko2013, Vonesh2012}). In particular the difference in interpretation in terms of population-level versus conditional effects have been both clarified and debated over the last decade. It has been argued that conditional modelling was to be preferred \citep{Lee2004}, but there seems to be pros and cons \citep[rejoinder]{Lee2004}. The conditional effects focus on a single subject, e.g. reporting a dose-response relationship with a subject-specific initial response at the lowest dose level. This initial effect might lead to a steeper curve, an earlier or later response, and a higher maximum response level compared to the population average. If no information about a specific individual is available to predict a response and just a general effect of a whole population of individuals is of interest, the conditional effects will not be an efficient way to summarize the population distribution. And it has to be noted that conditioning on the average subject effects will not result in the expected population response for nonlinear models.

In an attempt to bridge the gap between conditional and marginal models some authors have considered marginalization (e.g.,\citet{Heagerty2000, Wang2004, Parzen2011, Griswold2013}). However, these authors only considered special cases of generalized linear mixed-effects models, using non-standard distributional assumptions for the random effects or approximate attenuation factors to address the integration over the random effects distributions.

Conditional inference requires distributional assumptions at the level of each cluster next to assumptions on the distribution of each measurement whereas marginal inference may be carried out while completely ignoring the hierarchical structure. The marginal approach results in regression coefficients that can be interpreted as marginal effects but their standard errors will not fully incorporate the correlation between observations within individuals; i.e. for generalized estimation equation models \citet{Liang1986} showed that the standard errors of estimated marginal regression coefficients may be consistently estimated, even when the assumed within-cluster correlation is incorrect. The relative efficiency of the estimator will, however, be substantially increased by using a correctly specified correlation structure, especially in case of highly correlated coefficients.

For clustered data with nonlinear trends the arguments in favour of conditional modelling using nonlinear mixed-effects methodology are even more compelling than in the case for non-normal clustered data. In the context of generalized linear mixed-effects models all parameters are defined on the same scale introduced through the link function, which establishes the connection between the data scale and the so-called model scale of the linear predictor \citep{Stroup2014}. In contrast, for nonlinear clustered data usually several model scales are needed to accommodate all model parameters entering the nonlinear model equation. For instance, three model scales, asymptotes, steepness, and ED$_{50}$, are in use for the four-parameter log-logistic model, which we will introduce in detail below. 
Random effects are ideally suited to describe varying magnitudes of variation inherent to these different scales and different random effects on different scales will ensure that variation is captured in terms of the scale where it is present. It is difficult to imagine that marginal models can achieve a similar description of the variation based on the data scale only. Moreover, as pointed out by \citet{Griswold2013} choosing a conditional modelling approach does not rule out considering marginal interpretations of parameter estimates of interest. Indeed, conditional modelling allows both conditional and marginal interpretations of results. It is noteworthy that choosing a marginal modelling strategy does not provide the same flexibility as only marginal interpretations will eventually be possible \citep[p.~246]{McCulloch2008}. 

Instead of the direct estimation of marginal parameters as in \citet{Heagerty2000}, we focus on calculating marginal predictions by means of population averages. Whereas the random effects in a nonlinear hierarchical model can be defined on the different scales of the model parameters, estimating nonlinear model coefficients conditional on these random effects, the main interest might lie on the population averaged predictions. Based on the marginal predictions related marginalized parameters can be derived.

In the present study, the marginalization approach, which was recently proposed by \citet{Gerhard2014} for a specific nonlinear model for analysis of quantitative real-time polymerase chain reaction (QPCR) data, is extended for arbitrary nonlinear mixed-effects regression models, i.e., allowing for arbitrary random effects structures and also correlated residual errors. This extension leads a novel and operational combination of numerical integration based on Gaussian quadrature and the delta method for error propagation. 

We will focus on important applications in pharmacology and toxicology where dose-response models play a prominent role in evaluation of toxicity of chemicals. Such assessments usually involve one or more quantities derived from the model parameters, e.g., effective doses that correspond to a 10\%, 20\%, and 50\% reduction in toxicity (ED$_{10}$, ED$_{20}$, and ED$_{50}$). These parameters are interpreted by toxicologists naturally in a population context (e.g., \citealp{Weimer2012}) but unfortunately the estimates that are currently available when fitting hierarchical models do not allow for such an interpretation. Hence, there is an urgent need for a flexible marginalization approach that is adapted to the more complex hierarchical designs that now also becoming more common in toxicological experiments. Specifically, we extend the comprehensive infrastructure for dose-response modelling outlined by \citet{Ritz2005}, but only for models without random effects, to the case of dose-response mixed-effects models. 

We demonstrate its usefulness through two case studies exhibiting different aspects of clustered data: a toxicity evaluation in terms of effective doses in a human toxicology study that was replicated six times over time, and an assessment of potency in a field trial comparing two pesticides using five blocks.

Section~2 introduces nonlinear mixed-effects models in general. In Section~3 the marginalization approach is derived. A short simulation study is presented in Section~4. Section~5 report results from two case studies. Finally, we offer a perspective on the developed methodology in Section~6.

\section{Hierarchical nonlinear regression models}

Following the notation of \citet{Davidian2003}, a nonlinear regression model with a single hierarchical level can be defined in two stages that parametrise the variation within and between specific curves, respectively. 

\noindent \emph{Stage~1:} For the $i$th individual ($i=1,\dots,m$), we assume the following nonlinear regression model:
\begin{equation} \label{stage1} 
y_{ij} = f(x_{ij}, \boldsymbol{\beta}_{i}) + \epsilon_{ij} 
\end{equation}
where $\{y_{i1}, \ldots, y_{i n_i} \}$ and $\{x_{i1}, \ldots, x_{i n_i} \}$ denote the vectors of response values and covariate levels, respectively. A population of $i$ different curves is characterized by the nonlinear function $f$ through the curve-specific effect $\boldsymbol{\beta}_{i}$ (a $q \times 1$ vector). The within-cluster variation is explained through the residual error vector $\epsilon_{i} = (\epsilon_{i1}, \ldots, \epsilon_{in_i})$, which is assumed to be mean-zero normally distributed with variance-covariance matrix 
$\boldsymbol{\Lambda}_{i}$. In practice this matrix is often structured as $\sigma^2 I_{n_i}$ or $diag(\sigma_1^2, \ldots, \sigma_{n_i}^2)$. 

\noindent \emph{Stage~2:} The between curve variation is captured by splitting the curve-specific effect $\boldsymbol{\beta}_{i}$ into components describing the systematic and random variation between curves:
\[ \boldsymbol{\beta}_{i} = \boldsymbol{A}_{i}\boldsymbol{\beta} + \boldsymbol{B}_{i}\boldsymbol{b}_{i} \]
where $\boldsymbol{A}_{i}$ and $\boldsymbol{B}_{i}$ denote the fixed effects and random effects design matrices, respectively, $\boldsymbol{\beta}$ denotes the fixed effects parameters (a $p \times 1$ vector with $p \le q$), and the $b_i$'s denote the curve-specific random effects. The random effects can be assumed to follow a mean-zero normal distribution with a variance-covariance matrix denoted $\boldsymbol{G}$, which usually is simply the unstructured matrix. In what follows we only consider models with dummy coded design matrices for categorical predictors. These hierarchical models will still cover many experimental settings used in practice, such as comparison of several treatments in a dose-response experiment.

\section{Quadrature}

As already mentioned in the introduction the fixed effect parameters in nonlinear mixed-effects models usually cannot in general be interpreted as population quantities that are independent of the random effects, due to the inequality:
\begin{equation} \label{margcond} 
f(x_{ij}, \boldsymbol{A}_{i}\boldsymbol{\beta} + 
\boldsymbol{0}) \neq E\left\{f(x_{ij}, 
\boldsymbol{\beta}_{i}) \right\} = \tilde{f}(x_{j}, 
\tilde{\boldsymbol{A}_{i}} \tilde{\boldsymbol{\beta}}) 
\end{equation}
where $\tilde{f}$ denotes the unknown marginal mean function and, in particular, $\tilde{\boldsymbol{\beta}}$ denotes the parameters we are interested in. 
In contrast, $f$ denotes the mean function that was specified as part of the model specification. At this point we should mention that there are a few exceptions such as applications of nonlinear mixed-effects models with additive random effects on the data scale, meaning that marginal effects are directly available (e.g., \citet{Blankenship2003}, \citet[Chap. 6]{Demidenko2013}). 

The marginal expectation for a nonlinear mixed-effects model is calculated as follows:
\[ \mathrm{E}\left\{f(x_{ij}, \boldsymbol{\beta}_{i}) \right\} = \int \cdots \int f \{x_{ij}, (\beta_{i1} + b_{1}, \dots, \beta_{ip} + b_{p})\} \; \Phi(b_{1}, \dots, b_{p},  \boldsymbol{G}) \; d b_{1} \cdots db_{p} \]
where $\Phi()$ denotes the density function of a $p$-dimensional normal distribution with mean vector $\boldsymbol{0}$ and variance-covariance matrix $\boldsymbol{G}$.  Note that different nonlinear mixed-effects models will lead to  slightly different definitions of population-average effects. Specifically, this implies that derived marginal effects are affected by the distributional assumptions made for the random effects. 

Through a change of variables the original random effects may be transformed into a vector of independent and identically distributed random effects: 
$\boldsymbol{b}_{i}=\boldsymbol{\Omega u}_{i}$  where $\boldsymbol{G} = \boldsymbol{\Omega\Omega}^T$ is based on the Cholesky decomposition and $\boldsymbol{u}_{i} \sim N(\boldsymbol{0}, \boldsymbol{I}_{p \times p})$. We may then approximate the integral by the weighted sum:
\begin{equation} \label{approx}
\mathrm{E}\left\{ f(x_{ij}, \boldsymbol{\beta}_{i}) \right\} \approx \sum_{n=1}^{N} w_{n} f\{x_{ij}, (\beta_{i1} + \xi_{1n}, \dots, \beta_{ip} + 
\xi_{pn})\}, \quad \mathrm{with} \quad w_{n} = \prod_{r=1}^{p} w_{rn}, 
\end{equation}
using numerical Gauss-Hermite quadrature assuming a $p \times N$ matrix of nodes $\boldsymbol{\Psi}$ and corresponding weights $w_{rn}$, based on independent standard normal distribution functions \citep{Smyth1998a}. The quadrature nodes are transformed by $\boldsymbol{\xi} = \boldsymbol{\Psi^T \widehat{\Omega}}$ to reflect the scales and covariances of the random effects. The approximation to the integral can be thought of as an average of the individual curves, but instead of considering a simple mean involving the predicted random effects (the so-called EBLUPs), the average is based on the estimated multivariate normal distribution of the random effects. Standard errors of derived parameter estimates are obtained by applying the delta method to Eq.~(\ref{approx}) \citep[Chap. 5]{Vaart1998} using a numerically calculated gradient and extending the results by \citet{Gerhard2014}.

\subsection{Dose-response models}

We will focus on the model functions $f$ in Eq.~(\ref{stage1}) used for the analysis of dose-response data. A number of different model functions have been proposed in the literature; a short overview and references can be found in \citet{Ritz2010}. However, in practice only a few functions appear to be used routinely. Among these models the sigmoidal log-logistic model function is by far the one used the most. We define the five-parameter log-logistic model function as follows:
\begin{equation} \label{LL5}
f(x, \boldsymbol{\beta}) = \beta_2 + 
\frac{\beta_3-\beta_2}{(1+\exp[\beta_1\{\log(x)-\log(\beta_4)\}])^{\beta_5}} 
\end{equation}
The commonly used four-parameter version is obtained by setting $\beta_5=1$; this model is routinely applied in biology, toxicology and weed science (e.g., \citet{Price2012}). One reason for the popularity of this model is the fact that the model parameters have direct biological interpretations: The parameters $\beta_2$ and $\beta_3$ denote the lower and upper horizontal asymptotes that confine the sigmoidal regression curve. The parameter $\beta_4$ denotes the dose resulting in a halfway reduction between the limits $\beta_2$ and $\beta_3$ (which corresponds to the ED$_{50}$ in mixed-effects dose-response models, which are symmetrical increasing/decreasing around an inflection point, but might be influenced by parameters for asymmetry or hormesis). The parameter $\beta_1$ indicates the steepness of the curve in its transition between the two horizontal asymptotes. The fifth parameter $\beta_5$ is an asymmetry parameter leading to asymmetry in one or the other tail for $\beta_5 <1$ and $\beta_5 >1$, respectively.

Many a time not only the model parameters explicitly included in the model equation are of interest, but also some derived parameters, like the effective dose resulting in a $100\alpha\%$ ($0 < \alpha < 1$) reduction relative to the lower and upper limits (i.e., ED$_{100\alpha}$) or the ratio of two such effective doses (often referred to as the relative potency). Inference for derived parameters may be based on using the delta method, which results in an estimated approximate (asymptotically-based) variance-covariance matrix of the derived parameters. 

For a given fixed effects parameter configuration $\boldsymbol{A}_{0} \boldsymbol{\beta}$ (e.g., a specific treatment) the effective dose ED$_{100\alpha}$ is defined as the solution to the following inverse regression problem:
\[ \tilde{f}(ED_{100\alpha}, \boldsymbol{A}_{0} \boldsymbol{\beta}) = \alpha \tilde{f}(\infty, \boldsymbol{A}_{0} \boldsymbol{\beta}) + (1 - \alpha) \tilde{f}(0, \boldsymbol{A}_{0} \boldsymbol{\beta}) \]
which may be re-arranged into:
\begin{equation} \label{EDval}
\alpha = \frac{\tilde{f}(ED_{100\alpha}, \boldsymbol{A}_{0} \boldsymbol{\beta}) - \tilde{f}(0, \boldsymbol{A}_{0} \boldsymbol{\beta})}{\tilde{f}(\infty, \boldsymbol{A}_{0} \boldsymbol{\beta}) - \tilde{f}(0, \boldsymbol{A}_{0} \boldsymbol{\beta})}
\end{equation}
By definition ED$_{100\alpha}$ values are relative quantities that are defined relative to the lower and upper limits $\beta_2$ and $\beta_3$ in Eq.~(\ref{LL5}), which corresponds to $\tilde{f}(0, \boldsymbol{A}_{0} \boldsymbol{\beta})$ and $\tilde{f}(\infty, \boldsymbol{A}_{0} \boldsymbol{\beta})$, 
respectively. Instead of the population-level effective dose, a conditional ED$_{100\alpha}$ may also be calculated from the inverse regression of $f(ED_{100\alpha}, \boldsymbol{A}_{0} \boldsymbol{\beta} + \boldsymbol{b}_{0})$, that is conditioning on a given known random effects vector $\boldsymbol{b}_{0}$, which is often taken to be $\boldsymbol{0}$. However, both types of effective doses may be derived using the same nonlinear mixed-effects model. 

A related approach used for summarizing dose-response data is the benchmark dose methodology \citep{Piegorsch2010}. \citet{Ritz2013} proposed an operational definition of the benchmark dose for continuous endpoints, allowing for the incorporation of an a priori specified background level $p_0$ and risk level (benchmark response: $BMR$). This definition allows benchmark doses to be estimated as certain effective doses.

\subsection{Software implementation}
\label{software}

Our approach is implemented in {\tt{R}} \citep{Rlang2014}. Specifically, we developed the extension package {\tt{medrc}}, which combines the dose-response modelling framework of the package {\tt{drc}}, which allows nonlinear least-squares estimation for one or more dose-response curves through a convenient infrastructure for model specification and automatic calculation of suitable starting values \citep{Ritz2005}, with nonlinear mixed-effects estimation using the package {\tt{nlme}}  \citep{Pinheiro2000}.

This framework offers several key features such as having a set of predefined dose-response functions directly available, providing a unified 
parametrisation  (\texttt{b}: steepness, \texttt{c,d}: lower and upper asymptotes, \texttt{e}: inflection point location) and an easy-to-use formula interface for specifying the dose-response relationship and for parametrisation of modelling multiple curves and random effects. 

The package {\tt{medrc}} is available on GitHub under \\ \url{https://github.com/daniel-gerhard/medrc.git}. An accompanying vignette shows the full functionality of the package through additional examples.

\section{Simulation Study}

The accuracy of the proposed marginalization approach was investigated through simulations and comparisons to the marginal estimates obtained by fitting a generalized nonlinear least squares (GNLS) model assuming a compound-symmetry structure and an ordinary nonlinear least-squares regression (NLS) model, which entirely ignores the between-cluster variability. Both these marginal models provide parameter estimates that can be directly interpreted as population parameters, but the information about cluster effects is lost, i.e., the information is not quantified and therefore not incorporated as a separate contribution in the estimated standard errors, if appropriate. We also compared 
the marginalized estimates to the conditional estimates obtained directly from the nonlinear mixed effects model fits. As the conditional parameters have a different interpretation and the marginal parameters are even derived from different models, it is not expected that these approaches will result in estimates for the marginalized parameter. But it might be possible to characterise the performance of the marginalized estimates relative to these alternatives.

The simulation scenario was similar to the first case study introduced in the next section: We simulated from a three-parameter log-logistic model with a random intercept on each of the three parameters and a dose range from 0.01 to 3 with 10 equally spaced dose levels on the log-scale. The number of clusters was varied from 2 to 20 to evaluate the precision of the derived parameter estimates as the standard deviations of the random effects become more precisely determined. The specific configuration of fixed effects parameters and standard deviations is shown in Table~\ref{table1}. Note the very different magnitudes of the variation on the three model scales. In order to investigate the influence of different covariance structures of the random effects and corresponding model misspecification, either a diagonal or unstructured covariance matrix was assumed for the true and modelled dependence structure. 

\begin{table}[htb]
\caption{Parameter settings in the simulation study.}
\begin{center}
\begin{tabular}{lrrcrrrcrrr}
\hline
& Fixed   & Random effects & & \multicolumn{7}{c}{Correlation of random effects} 
\\
Parameter   & effect & Std. Dev.   &  & \multicolumn{3}{c}{Unstructured}  & & 
\multicolumn{3}{c}{Diagonal} 
\\
\hline
Steepness ($\beta_1$) & 5    & 0.5  & & 1    & -0.9 & 0.8        & & 1 & 0 & 0 
\\
Upper asymptote ($\beta_2$)  & 2000 & 500 & & -0.9 & 1    & -0.5 & & 0 & 1 & 0 
\\
ED$_{50}$ ($\beta_4$) & 0.5  & 0.1 & & 0.8  & -0.5 & 1 & & 0 & 0 & 1 \\
Residuals ($\epsilon_{ij}$)  &      & 100 \\
\hline
\end{tabular}
\end{center}
\label{table1}
\end{table}

\begin{figure}[htb]
\begin{center}
\includegraphics[width=16cm]{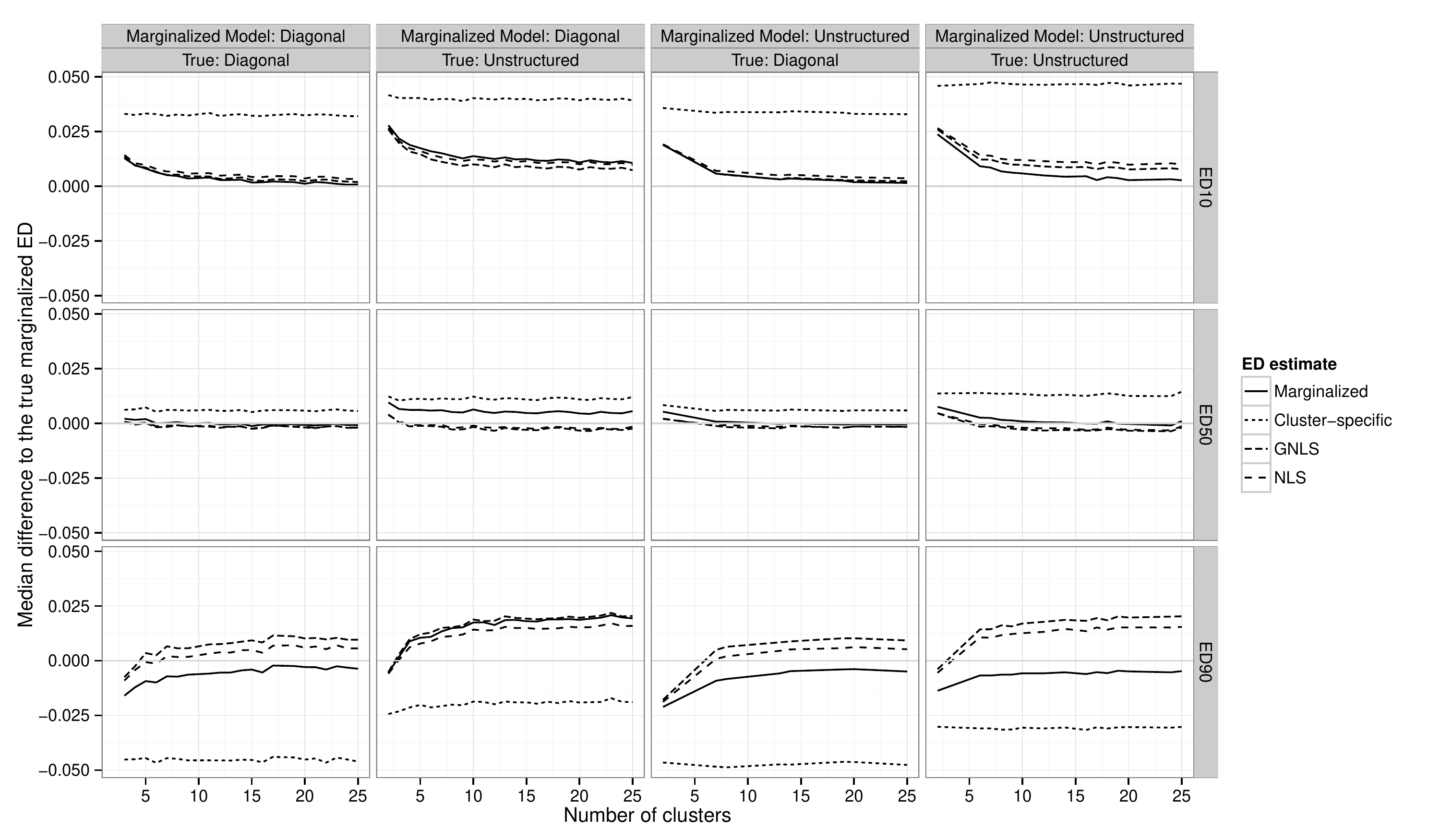}
\end{center}
\caption{Simulation result: Median of estimated differences to the true marginalized effective dose for marginalized, conditional, GNLS, and NLS estimates of ED$_{10}$, ED$_{50}$, and ED$_{90}$ (rows) using either an unstructured or diagonal variance-covariance matrix for the true and/or modelled random effects (columns). The true parameter corresponds to the dashed horizontal line at 0.}
\label{figure1}
\end{figure}

The `true' marginal effective dose parameters were obtained by Monte Carlo integration with 100,000 sampling steps for each parameter settings; this means sampling directly from a multivariate normal distribution conditioning on the pre-defined, true covariance of the random effects. The accuracy of the effective doses was evaluated by the median deviation of the estimated parameter relative to the true marginal parameter as shown in Figure~\ref{figure1}. This `bias' has to be cautiously interpreted for the conditional and marginal approaches, as it shows the deviation to the true marginalized effective dose, which of course has a different interpretation.

The marginalized estimates had the smallest deviation to the `true' effective dose in every scenario with a correctly specified correlation structure. The conditional estimate instead remained constant regardless of the number of clusters. Only when the model was misspecified, assuming a diagonal instead of a true unstructured covariance matrix of the random effects, the marginalized estimates showed an increased deviation even at a large number of clusters, but nevertheless this deviation is still in a similar range as for the marginal estimates.

By increasing the standard deviation of the random effects associated with the parameter ED$_{50}$ tenfold, the differences between conditional and marginalized estimated effective doses increased, resulting in an increasing difference for the conditional estimates (results are shown as Supporting Information). The differences for the conditional and marginalized estimates were always going in the same direction. For ED$_{50}$ and even more so for ED$_{10}$ the conditional estimates showed a positive difference whereas there was a negative difference for ED$_{90}$; this finding is a result of assuming that random effects are associated with the upper asymptote but not with the lower asymptote, which was fixed at 0.

It is worth noting that overall the two marginal approaches yielded very similar results even though they involved different assumptions about the correlation structure. By generating the data given a diagonal random effects structure smaller deviation was observed, as it can be better approximated by the compound symmetry assumption on the residuals than the unstructured covariance. Moreover, for ED$_{10}$ and ED$_{50}$ the marginal estimates behaved like the marginalized estimates, but for ED$_{90}$ they resulted in a deviation in the opposite direction from the conditional and marginalized estimates.

\begin{figure}[htb]
\begin{center}
\includegraphics[width=16cm]{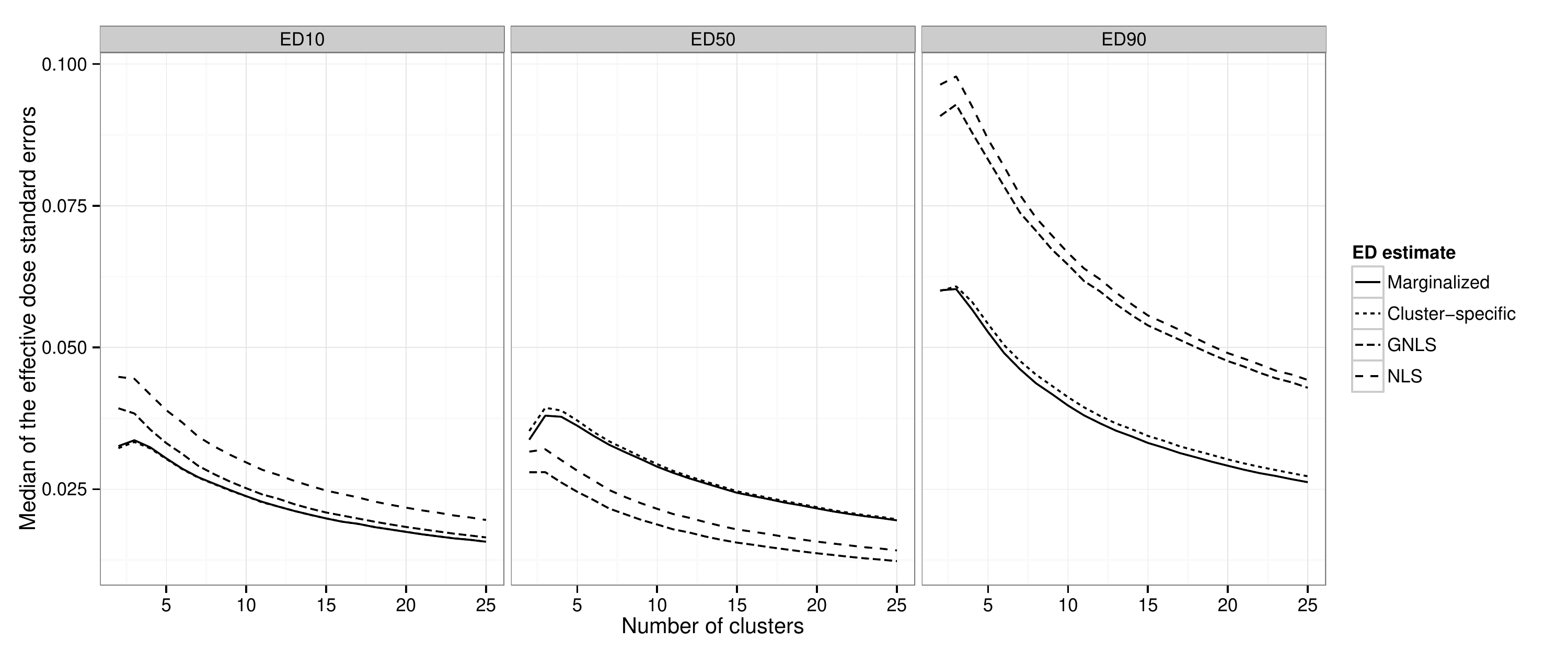}
\end{center}
\caption{Simulation result: Median of estimated standard errors for marginalized, conditional, GNLS, and NLS estimates of ED$_{10}$, ED$_{50}$, and ED$_{90}$. An unstructured variance-covariance matrix of the random effects was used for data generation and modelling.}
\label{figure2}
\end{figure}

Figure~\ref{figure2} shows the medians of the estimated standard errors of the effective dose estimates from the simulation study. The standard errors of the marginalized estimates are comparable to the ones of the conditional estimates. The dependence of the marginal estimates on the different scales of the random effects is apparent: In comparison to the proposed marginalization approach the two marginal models, which showed very similar behaviour, resulted in estimated standard errors for the estimated ED$_{50}$ that were somewhat smaller, but became larger with increasing distance from the inflection point. For the estimated ED$_{90}$ (near the lower asymptote that was fixed at zero) the marginal estimates had standard errors that were larger by a factor 1.5--2, indicating the inability of the marginal models to borrow strength across different model scales. 

\section{Case Studies}

The following case studies were analysed throughout using the {\tt{R}} packages {\tt{drc}} and {\tt{medrc}} as previously described. Source code to reproduce the results is available in the documentation of the package {\tt{medrc}} (see Section \ref{software}).

\subsection{An example with between-assay variation}

\citet{Nellemann2003} carried out experiments to assess the in vitro effects of the fungicide vinclozolin. The data were obtained using an androgen receptor reporter gene assay, which was repeated six times (on different days). Each assay resulted in concentration-response data with nine concentrations (in $\mu$M) of the fungicide, and the response measured was chemiluminescence (in luminescence units). The same nine concentrations were used in all six assays. However, in one assay results were only obtained for eight concentrations. 

A plausible biological assumption was that no signal would be observed for very high fungicide concentrations, indicating that a three-parameter log-logistic model with a lower asymptote at 0 ((Eq.~(\ref{LL5}) with $\beta_2=0$ and $\beta_5=1$)) would be a realistic model for the data. Assay-specific random effects on each of the three model parameters, but not on the fixed lower asymptote. This model allowed summarizing the between-assay variation in terms of a $3 \times 3$ variance-covariance matrix.

A marginal generalized nonlinear least-squares model (GNLS) is used for comparison. Instead of a full random effects representation of the between-assay variability on the scales of the fixed effects parameters, a compound-symmetry structure is assumed for the within-assay residuals.

\begin{figure}[htb]
\begin{center}
  \includegraphics[width=10cm]{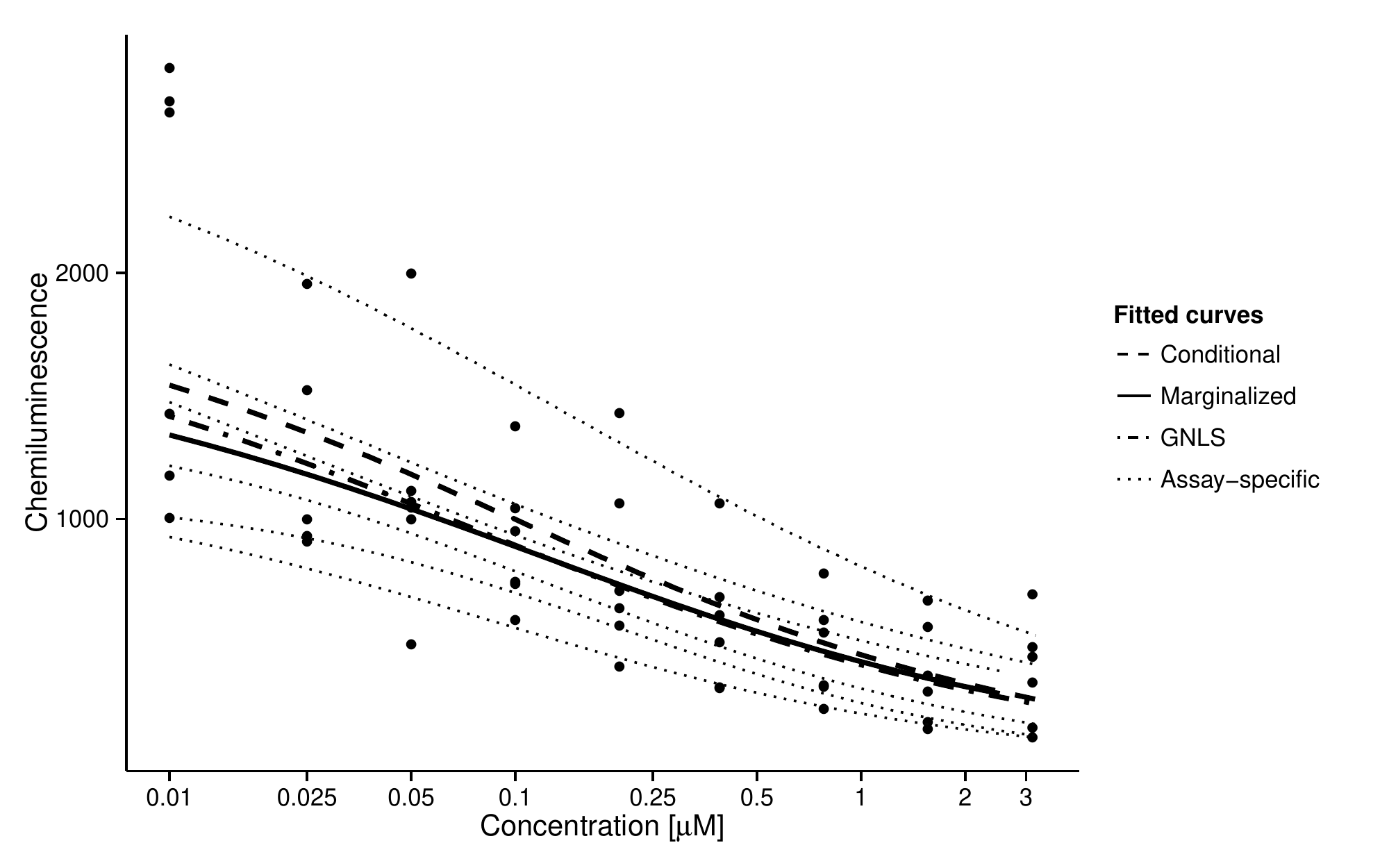}
\end{center}
\caption{Chemiluminescence observations together with conditional, marginalized, and assay-specific fitted curves based on a nonlinear 
mixed-effects regression model assuming a three-parameter log-logistic regression relationship. The corresponding marginal prediction based on GNLS is also shown.}
\label{figure3}
\end{figure}

The fitted dose-response curves for the conditional, marginal, and marginalized approaches are shown in Figure \ref{figure3} along with the individual assay-specific fitted curves (inserting estimated random effects). In this case they agreed quite well for high concentration but the marginalized fit is below the conditional fit and slightly below the marginal fit for low concentrations where the response is more variable. Estimated effective doses are shown in Table~\ref{table2}. For large concentrations ($>$ ED$_{50}$) the standard errors were smaller for the marginalized approach than for the marginal and only slightly larger than for the conditional approach, indicating that the gain in precision achieved using a nonlinear mixed-effects model is retained through marginalization. For the highest ED levels the standard errors in the marginal approach were twice as large compared to the conditional and marginalized approaches. This is presumably the result of modelling the within-assay correlation on the scale of the residuals, independent of the difference in scales of the two asymptotes.

\begin{table}[htb]
\caption{Effective dose estimates for the vinclozolin study obtained from 
conditional, marginal (GNLS), and marginalized models.}
\begin{center}
\begin{tabular}{lrrrrrr}
\hline
  & \multicolumn{2}{c}{Conditional} & \multicolumn{2}{c}{Marginalized} & 
\multicolumn{2}{c}{Marginal (GNLS)} \\
ED & Estimate & Std.Error             & Estimate & Std.Error         & Estimate 
& Std.Error \\
\hline
10 & 0.002    & 0.001                 & 0.001    & 0.001             & 0.001    & 0.001 \\
25 & 0.013    & 0.007                 & 0.010    & 0.007             & 0.009    & 0.004 \\
50 & 0.101    & 0.031                 & 0.097    & 0.039             & 0.075    & 0.028 \\         
75 & 0.781    & 0.141                 & 0.834    & 0.196             & 0.655    & 0.393 \\ 
90 & 6.012    & 1.949                 & 7.638    & 2.666             & 5.710    & 5.394 \\
\hline
\end{tabular}
\end{center}
\label{table2}
\end{table}

\subsection{An example with between-assay variation and different treatments}

\citet{Streibig1999} investigated the inhibition of photosynthesis in response to two synthetic photosystem II inhibitors, the herbicides diuron and bentazon. In an experiment, the effect of oxygen consumption of thylakoid membranes (chloroplasts) from spinach was measured after incubation with the synthetic inhibitors. Five assays, three treated with bentazon and two with diuron, were used. For each assay six increasing herbicide concentrations were applied together with a negative control, using different dose ranges on a logarithmic scale for the two treatments to encompass the whole dose-response range based on preliminary experiments.

For the comparison of the two herbicides, dose-response curves were fitted assuming a four-parameter log-logistic model with a separate set of fixed effects coefficients for each of the two treatments (in total 8 model parameters). Random effects were included for each of the four parameters to model the between-assay variability in the different model scales. Using the information about the between-assay variability is especially advantageous as the dose levels for the two herbicides did not cover the same dose range. The two fixed effect curves were compared by means of relative potencies based on ED$_{10}$, ED$_{20}$, ED$_{50}$, ED$_{75}$, ED$_{90}$ estimates. 

Again, a generalized nonlinear least-squares model (GNLS) is used for comparison, assuming a compound-symmetry structure of the residuals to reflect the within-assay correlation. 

\begin{figure}[htb]
\begin{center}
  \includegraphics[width=10cm]{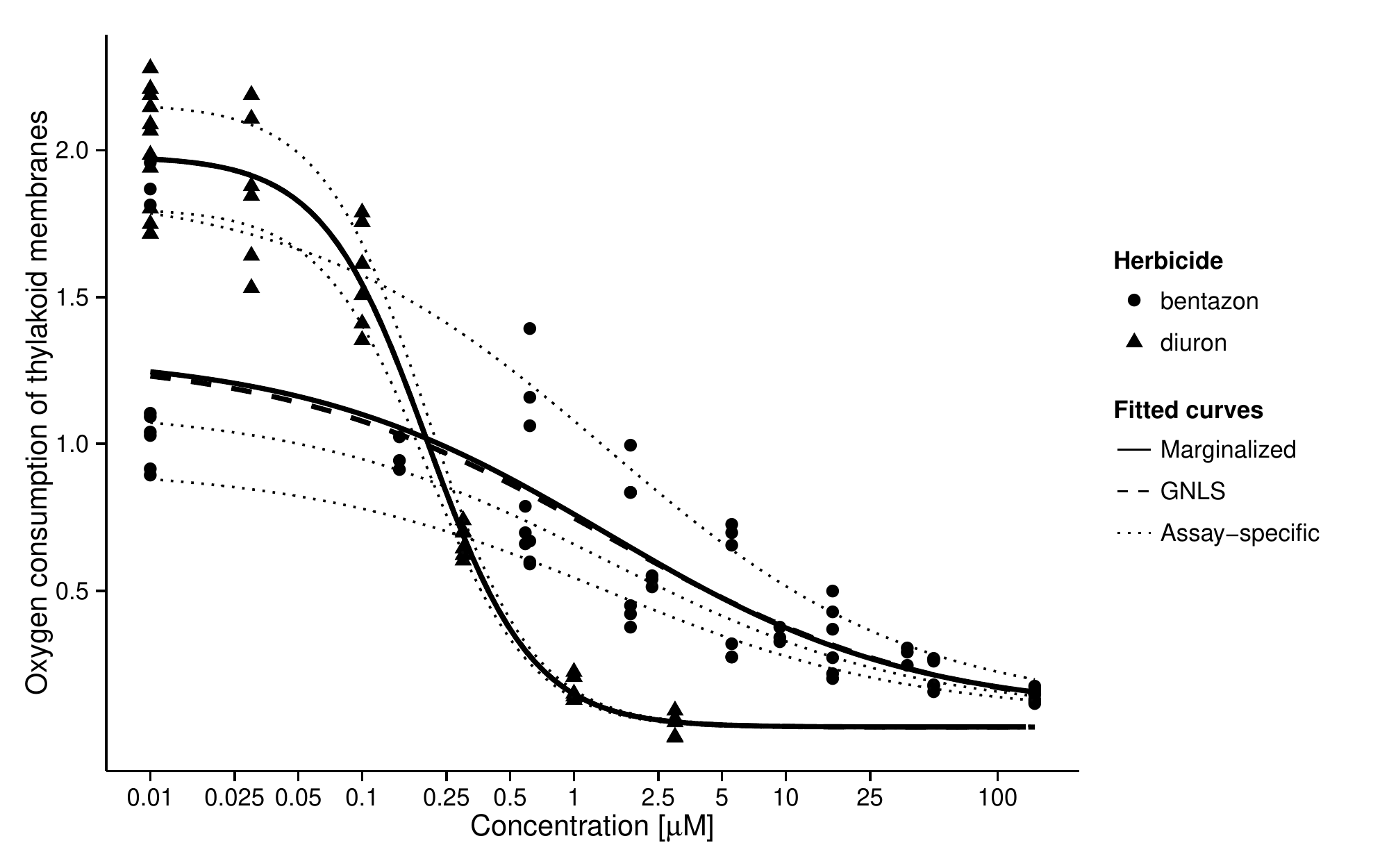}
\end{center}
\caption{Observations of oxygen consumption and conditional, marginalized, and assay-specific fitted curves based on a nonlinear mixed-effects regression model assuming a four-parameter log-logistic regression relationship for each treatment group. The corresponding marginal prediction based on GNLS is also shown. For the herbicide treatments their corresponding assay-specific curves are shown, predicting a response at each assay level.}
\label{figure4}
\end{figure}

The fitted mean dose-response curves, derived using the marginal (GNLS) and marginalized approaches, as well as the assay-specific fitted curves are shown in Figure~\ref{figure4}. We can conclude that the marginalized approach recovers the marginal mean curve very accurately in this case as the fitted mean curves are all very similar. The estimated relative potencies are shown in Table~\ref{table3}. The conditional and marginalized estimates were identical and this may be explained by the fact that the between-assay standard deviation for the slope and ED$_{50}$ parameters ($\beta_1$ and $\beta_4$) effectively were equal to 0; with random effects only on the scales on the asymptotes of a four-parameter log-logistic model, the fixed effects can be interpreted as marginal effects. For the marginal model the estimated within-assay correlation coefficient for the compound symmetry structure was 0.55, but it is not specific to the different scales of the two asymptotes. Therefore, the confidence intervals were narrower for small ED values and wider for the high ED values.

\begin{table}[htb]
\caption{Estimated effective doses and corresponding relative potencies for the spinach data example.}
\begin{center}
\begin{tabular}{llrrcrr}
\hline
		&    &  \multicolumn{2}{c}{Effective Dose}    &  Relative   &  
\multicolumn{2}{c}{Confidence limits} \\
Model            & ED & Bentazon & Diuron & potency & Lower & Upper \\
\hline
		& 10 & 0.038    & 0.058  & 0.652   &   0.280  &   1.056 \\
		& 25 & 0.232    & 0.108  & 2.144   &   1.420  &   2.911 \\
Conditional     & 50 & 1.433    & 0.203  & 7.056   &   5.217  &   8.980 \\
		& 75 & 8.832    & 0.380  & 23.220  &  13.112  &   33.940 \\
		& 90 & 54.437   & 0.712  & 76.415  &  23.776  &  133.970 \\
\hline
		& 10 & 0.038    & 0.058  & 0.651   & 0.278    &   1.058   \\
		& 25 & 0.232    & 0.108  & 2.144   & 1.421    &   2.911   \\
Marginalized    & 50 & 1.433    & 0.203  & 7.056   & 5.216    &  8.980   \\
		& 75 & 8.832    & 0.380  & 23.220  & 13.112   &  33.940    \\
		& 90 & 54.437   & 0.712  & 76.415  &  23.776  &  133.969   \\
\hline
		& 10 & 0.030    & 0.058  & 0.513   &  -0.147  &  1.278  \\
		& 25 & 0.207    & 0.108  & 1.909   &   0.584  &  3.382   \\
Marginal        & 50 & 1.445    & 0.203  & 7.110   &  2.385   &  12.157  \\
		& 75 & 10.076   & 0.381  & 26.478  &  -6.314  &  61.692    \\
		& 90 & 70.286   & 0.713  & 98.601  & -95.172  & 318.229   \\
\hline
\end{tabular}
\end{center}
\label{table3}
\end{table}

\section{Discussion}

We have proposed a novel and flexible marginalization approach for obtaining marginal estimates from nonlinear mixed-effects models that are routinely used in toxicology and biological science applications. The approach combines the best of two worlds: interpretable population-level estimates obtained using models that allow to borrow strength across observations. 

For all of the two case studies, the population average interpretation is useful, looking at a representative sample of subjects characterizing the population of possible subject effects. Basing the interpretation of a dose effect on a single subject will not be easy to generalise for a range of other future samples from the subject population; this is different for e.g. medical applications where the benefit of a single patient might be of foremost interest instead of a general population effect.

We specifically focused on nonlinear mixed-effects dose-response models although the proposed methodology is also applicable to other types of 
nonlinear mixed-effects regression models with normally distributed random effects.

As an alternative to the Gauss-Hermite quadrature, Monte Carlo integration could be used, sampling directly from a multivariate normal distribution conditioning on the estimated covariance of the random effects. But as it is a computationally very demanding approach we do not at present recommend it for routine analyses.

A limitation of the marginalization approach is the conditioning on estimated standard deviations of the random effects. This means we ignore the uncertainty involved in using estimated variance components, potentially leading to underestimation of the uncertainty of our parameter estimates of interest with consequences for derived p-values and confidence intervals.

\bibliographystyle{apalike}

\appendix

\label{lastpage}

\end{document}